\documentclass[12pt]{article}
\usepackage{amssymb,amsmath,epsfig}

\renewcommand{\theequation}{\arabic{section}.\arabic{equation}}

\begin{document}

\title{\bf Plasma Wave Properties of the Schwarzschild Magnetosphere in a Veselago Medium}
\author{M. Sharif \thanks{msharif@math.pu.edu.pk} and Noureen Mukhtar
\thanks{noureen.mukhtar@yahoo.com}\\
Department of Mathematics, University of the Punjab,\\
Quaid-e-Azam Campus, Lahore-54590, Pakistan.}
\date{}
\maketitle
\begin{abstract}
We re-formulate the $3+1$ GRMHD equations for the Schwarzschild
black hole in a Veselago medium. Linear perturbation in rotating
(non-magnetized and magnetized) plasma is introduced and their
Fourier analysis is considered. We discuss wave properties with the
help of wave vector, refractive index and change in refractive index
in the form of graphs. It is concluded that some waves move away
from the event horizon in this unusual medium. We conclude that for
the rotating non-magnetized plasma, our results confirm the presence
of Veselago medium while the rotating magnetized plasma does not
provide any evidence for this medium.
\end{abstract}
{\bf Keywords:} Veselago medium; $3+1$ formalism; GRMHD equations;
Isothermal plasma; Dispersion relations.\\
{\bf PACS:} 95.30.Sf; 95.30.Qd; 04.30.Nk

\section{Introduction}

Schwarzschild black hole is the simplest one which has neither
angular momentum nor charge. Plasma is the most common form of
matter in the universe. It can be accelerated and directed by
electromagnetic fields which make it to be controlled and
employed. The matter in inner-stellar space is composed of plasma.
The general relativistic magnetohydrodynamics (GRMHD) is the
standard theory which is helpful to explore the dynamics of
falling magnetized plasma towards the event horizon.

An accretion disk is formed near the Schwarzschild black hole as
plasma flows towards it. Perturbations in black hole regime have
always been fascinated the researchers. Regge and Wheeler \cite{1}
discussed the stability of Schwarzschild singularity. Price
\cite{2} examined the dynamics of non-spherical perturbations
during the collapse of stars with a scalar-field analog. Waves in
an electron-positron plasma were investigated by Sakai and Kawata
\cite{3} in a frame of two-fluid equation for the Schwarzschild
black hole. Southwood \cite{4} used a small perturbation approach
to derive the general stability criteria for plasma with isotropic
pressure. Gleiser et al. \cite{5} studied even parity perturbation
of the Schwarzschild black hole up to second order.

Arnowitt, Deser and Misner (ADM) \cite{6} developed 3+1 formalism in
which spacetime is foliated into layers of three-dimensional
spacelike hypersurface, threaded by a timelike normal. Many
relativists \cite{7}-\cite{8} applied this technique to illustrate
different attributes of general relativity. Thorne and Macdonald
\cite{9}-\cite{10} evolved the electromagnetic fields of the black
hole theory. Holcomb and Tajima \cite{11}, Holcomb \cite{12} and
Dettmann et al. \cite{13} analyzed the production of waves in the
Friedmann universe. Buzzi et al. \cite{14} examined the wave
properties of two fluid plasma in the surroundings of Schwarzschild
event horizon. Zhang \cite{15} explicated the theory of stationary
symmetric GRMHD in black hole regime. The same author \cite{16} also
studied the behavior of perturbation of cold plasma in the
magnetosphere of Kerr black hole. Sharif and his collaborators
\cite{17}-\cite{20} determined the plasma wave properties by using
dispersion relations. They analyzed it for cold, isothermal and hot
plasmas.

Veselago medium or Double negative medium (DNG) is the most famous
class of metamaterials. This medium has simultaneously negative
electric permittivity and magnetic permeability. Negative phase
velocity (NPV) and negative refractive index are main features of
this medium. Ziolkowski \cite{21} analyzed electromagnetic wave
properties in DNG. Valanju et al. \cite{22} established the
well-known idea that wave refraction in DNG medium is always
positive but not homogenous. Ross et al. \cite{23} found that NPV
propagation develops at lower values of cosmological constant
$\Lambda$ in the vicinity of non-rotating black hole. Mackey and
Lakhtakia \cite{24} verified that presence of charge raises the
tendency of a rotating black hole to support the NPV propagation in
its ergosphere. In a paper given in a book \cite{25}, J li
generalized the concept of negative medium to acoustic waves. In a
recent paper, Veselago \cite{26} has explained how an
electromagnetic wave transfers energy, linear momentum and mass in a
negative refraction-medium.

In the present paper, we discuss isothermal plasma wave properties
of the Schwarzschild black hole in a Veselago medium. The format
of the paper is as follows: General line element and its formation
in the Schwarzschild planar analogue is given in Section
\textbf{2}. In Section \textbf{3}, perturbed and Fourier analyzed
forms of 3+1 GRMHD equations for isothermal plasma are presented.
Sections \textbf{4} and \textbf{5} contain the restricted $3+1$
GRMHD equations for rotating (non-magnetized and magnetized)
plasmas and discussion related to wave properties. Summary of the
results is given in Section \textbf{6}.

\section{ADM 3+1 Formalism and Schwarzschild Planar Analogue}

The general line element in \emph{ADM 3+1 formalism} is given by
\cite{16}
\begin{equation}\setcounter{equation}{1}\label{1}
ds^2=-\alpha^2dt^2+\gamma_{ij}(dx^i+\beta^idt)(dx^j+\beta^jdt),
\end{equation}
where the lapse function $\alpha$ is  the ratio of FIDO (fiducial
observer) proper time to universal time, i.e.,
$\alpha=\frac{d\tau}{dt},~\beta^i$ is the shift vector and
$\gamma_{ij}~(i,j=1,2,3)$ are the components of three dimensional
absolute space. FIDO is the natural observer associated with the
above line element. In Schwarzschild planar analogue, shift vector
vanishes due to zero angular momentum and $\gamma_{ii}=1$ represents
Euclidean 3 geometry. Thus it reduces to the form \cite{17}
\begin{equation}\label{2}
ds^2=-\alpha^2(z)dt^2+dx^2+dy^2+dz^2,
\end{equation}
where the directions $z,~x$ and $y$ are analogous to the
Schwarzschild coordinates $r,~\phi$ and $\theta$ respectively.

\section{3+1 GRMHD Equations for Isothermal Plasma in Veselago Medium}

The $3+1$ GRMHD equations in a Veselago medium for the plasma
present in general line element and the Schwarzschild planar
analogue (Eqs.(\ref{1}) and (\ref{2})) are given in Appendix A. In
the vicinity of the Schwarzschild black hole, equation of state for
isothermal plasma is \cite{16}
\begin{eqnarray}\setcounter{equation}{1}\label{3}
\mu=\frac{\rho+p}{\rho_0},
\end{eqnarray}
here $\rho_0,~\rho,~p$ and $\mu$ are the rest mass density, moving
mass density, pressure and specific enthalpy respectively. The
specific enthalpy $\mu$ is constant here. This equation indicates
that there is no exchange of energy between the plasma and the
magnetic field of the fluid. For isothermal plasma existing in
Schwarzschild magnetosphere, the 3+1 GRMHD equations $(A10)$-$(A14)$
take the form
\begin{eqnarray}
\label{4} &&\frac{\partial \textbf{B}}{\partial
t}=-\nabla\times(\alpha \textbf{V}\times \textbf{B}),\\
\label{5}&&\nabla.\textbf{B}=0,\\
\label{6} &&\frac{\partial (\rho+p) }{\partial t}+(\alpha
\textbf{V}.\nabla)(\rho+p)+(\rho+p)\gamma^2\textbf{V}.
\frac{\partial \textbf{V}}{\partial t}+(\rho+p)\gamma^2\nonumber\\
&&V.(\alpha\textbf{V}.\nabla)\textbf{V}
+(\rho+p) \nabla.(\alpha\textbf{V})=0,\\
\label{7}&&\left\{\left((\rho+p)\gamma^2+\frac{\textbf{B}^2}{4\pi}\right)\delta_{ij}
+(\rho+p)\gamma^4V_iV_j-\frac{1}{4\pi}B_iB_j\right\}
\left(\frac{1}{\alpha}\frac{\partial}{\partial
t}\right.\nonumber\\
&&\left.+\textbf{V}.\nabla\right)V^j
-\left(\frac{\textbf{B}^2}{4\pi}\delta_{ij}-\frac{1}{4\pi}B_iB_j\right)
V^j_{,k}V^k+(\rho+p)\gamma^2a_i+p_{,i}\nonumber\\
&&=\frac{1}{4\pi}(\textbf{V}\times\textbf{B})_i\nabla.(\textbf{V}\times\textbf{B})
-\frac{1}{8\pi\alpha^2}(\alpha\textbf{B})^2_{,i}+\frac{1}{4\pi\alpha}(\alpha
B_i)_{,j}B^j\nonumber\\&&-\frac{1}{4\pi\alpha}
[\textbf{B}\times\{\textbf{V}\times(\nabla\times(\alpha\textbf{V}\times\textbf{B}))\}]_i,\\\nonumber
\label{8} &&(\frac{1}{\alpha}\frac{\partial}{\partial
t}+\textbf{V}.\nabla)(\rho+p)\gamma^2-\frac{1}{\alpha}\frac
{\partial p}{\partial t}+2(\rho+p)\gamma^2(\textbf{V}.\textbf{a})
+(\rho+p)\nonumber\\
&&\gamma^2(\nabla.\textbf{V})
-\frac{1}{4\pi\alpha}\left.(\textbf{V}\times\textbf{B}).(\textbf{V}\times\frac
{\partial \textbf{B}}{\partial t}\right.)
-\frac{1}{4\pi\alpha}\left.(\textbf{V}\times\textbf{B}).(\textbf{B}\times\frac{\partial
\textbf{B}}{\partial
t}\right.)\nonumber\\&&+\frac{1}{4\pi\alpha}\left.(\nabla\times\alpha\textbf{B}\right.)=0.
\end{eqnarray}

We assume that plasma flows in $xz$-plane so that the velocity $\bf
V$ and magnetic field $\bf B$ experienced by FIDO are given as
\begin{eqnarray}\label{9}
&&\textbf{V}=V(z)\textbf{e}_x+u(z)\textbf{e}_z,\nonumber\\
&&\textbf{B}=B[\lambda(z)\textbf{e}_x+\textbf{e}_z],
\end{eqnarray}
where $B$ is an arbitrary constant. The relation among $\lambda,~u$
and $V$ is given by \cite{17}
\begin{equation}\label{a}
V=\frac{V^F}{\alpha}+\lambda u,
\end{equation}
where $V^F$ is a constant of integration.
$\gamma=\frac{1}{\sqrt{1-\textbf{V}^2}}$ is the Lorentz factor which
becomes
\begin{equation}\label{b}
\gamma=\frac{1}{\sqrt{1-u^2-V^2}}.
\end{equation}

The gravity of black hole perturbs the plasma flow. Inserting the
linear perturbation to density $\rho$, pressure $p$, velocity
$\textbf{V}$ and magnetic field $\textbf{B}$, we have
\begin{eqnarray}\label{10}
&&\rho=\rho^0+\delta\rho=\rho^0+\rho\widetilde{\rho},\quad
p=p^0+\delta p=p^0+p\widetilde{p},\nonumber\\
&&\textbf{V}=\textbf{V}^0+\delta\textbf{V}=\textbf{V}^0+\textbf{v},\quad
\textbf{B}=\textbf{B}^0+\delta\textbf{B}=\textbf{B}^0+B\textbf{b},
\end{eqnarray}
where unperturbed and linearly perturbed quantities are denoted by
$\rho^0,~p,~\textbf{V}^0$, $\textbf{B}^0$ and $\delta \rho$, $\delta
p$, $\delta\textbf{V}$, $\delta\textbf{B}$ respectively. The
quantities $\widetilde{\rho},~\widetilde{p}, v_x, v_z, b_x$ and
$b_z$ are dimensionless which we shall introduce for the perturbed
quantities
\begin{eqnarray}\label{11}
&&\tilde{\rho}=\tilde{\rho}(t,z),\quad \tilde{p}=\tilde{p}(t,z),\nonumber\\
&&\textbf{v}=\delta\textbf{V}=v_x(t,z)\textbf{e}_x
+v_z(t,z)\textbf{e}_z,\nonumber\\
&&\textbf{b}=\frac{\delta\textbf{B}}{B}=b_x(t,z)\textbf{e}_x
+b_z(t,z)\textbf{e}_z,
\end{eqnarray}
The perfect GRMHD equations (Eqs.(\ref{4})-(\ref{8})), after the
insertion of linear perturbation from Eq.(\ref{11}), become
\begin{eqnarray}\label{12}
&&\frac{\partial(\delta\textbf{B})}{\partial
t}=-\nabla\times(\alpha\textbf{v}\times\textbf{B})
-\nabla\times(\alpha\textbf{V}\times\delta\textbf{B}),\\\label{13}
&&\nabla.(\delta\textbf{B})=0,\\\label{14}
&&\frac{\partial(\delta\rho+\delta
p)}{\partial t}+(\alpha\textbf{V}.\nabla)(\delta\rho+\delta
p)+(\rho+p)\gamma^2\textbf{V}.\frac{\partial \textbf{v}}{\partial
t}-\alpha(\rho+p)\nonumber\\
&&(\textbf{v}.\nabla\ln u)+\alpha(\rho+p)(\nabla.\textbf{v})
+(\delta\rho+\delta p)(\nabla.\alpha\textbf{V})+(\delta\rho+\delta
p)\gamma^2\nonumber\\
&&\textbf{V}.(\alpha\textbf{V}.\nabla)\textbf{V}
+2(\rho+p)\gamma^2(\textbf{V}.\textbf{v})(\alpha\textbf{V}.\nabla)
\ln\gamma+(\rho+p)\gamma^2\nonumber\\
&&(\alpha\textbf{V}.\nabla\textbf{V}).\textbf{v}+(\rho+p)
\gamma^2\textbf{V}.(\alpha\textbf{V}.\nabla)\textbf{v}=0,
\end{eqnarray}
\begin{eqnarray}\label{15}
&&\left\{\left((\rho+p)\gamma^2+\frac{\textbf{B}^2}{4\pi}\right)\delta_{ij}
+(\rho+p)\gamma^4V_iV_j-\frac{1}{4\pi}B_iB_j\right\}\frac{1}{\alpha}\frac{\partial
v^j}{\partial t}\nonumber\\
&&+\frac{1}{4\pi}[\textbf{B}\times\{\textbf{V}\times\frac{1}{\alpha}\frac{\partial
(\delta\textbf{B})}{\partial
t}\}]_i+(\rho+p)\gamma^2v_{i,j}V^j+(\rho+p)\nonumber\\
&&\times\gamma^4V_iv_{j,k}V^jV^k
-\frac{1}{4\pi\alpha}\{(\alpha\delta B_i)_{,j}-(\alpha\delta
B_j)_{,i}\}B^j+(\delta p)_i\nonumber\\
&&=-\gamma^2\{(\delta\rho+\delta
p)+2(\rho+p)\gamma^2(\textbf{V}.\textbf{v})\}a_i+\frac{1}{4\pi\alpha}\{(\alpha
B_i)_{,j}\nonumber\\
&&-(\alpha B_ j)_{,i}\}\delta
B^j-(\rho+p)\gamma^4(v_iV^j+v^jV_i)V_{k,j}V^k-\gamma^2\{(\delta\rho+\delta
p)\nonumber\\
&&V^j+2(\rho+p)\gamma^2(\textbf{V}.\textbf{v})V^j
+(\rho+p)v^j\}V_{i,j}-\gamma^4V_i\{(\delta\rho+\delta p)V^j\nonumber\\
&&+4(\rho+p)\gamma^2(\textbf{V}.\textbf{v})V^j+(\rho+p)v^j\}V_{j,k}V^k,\\\label{16}
&&\gamma^2\frac{1}{\alpha}\frac{\partial(\delta\rho+\delta
p)}{\partial
t}+\textbf{v}.\nabla(\rho+p)\gamma^2-\frac{1}{\alpha}\frac{\partial(\delta
p)}{\partial t}+(\textbf{V}.\nabla)(\delta\rho+\delta
p)\gamma^2\nonumber\\
&&+2(\rho+p)\gamma^4(\textbf{V}.\nabla)(\textbf{V}.\textbf{v})
+2(\rho+p)\gamma^2(\textbf{v}.\textbf{a})
+4(\rho+p)\gamma^4(\textbf{V}.\textbf{v})\nonumber\\
&&(\textbf{V}.\textbf{a})+2(\delta\rho+\delta
p)\gamma^2(\textbf{V}.\textbf{a})+(\rho+p)\gamma^2(\nabla.\textbf{v})+2(\rho+p)\gamma^4\nonumber\\
&&(\textbf{V}.\textbf{v})(\nabla.\textbf{V})+(\delta\rho+\delta
p)\gamma^2(\nabla.\textbf{V})=\frac{1}{4\pi\alpha}
[\textbf{v}.(\textbf{B}.\frac{\partial\textbf{B}}{\partial
t})\textbf{V}+\textbf{V}.(\textbf{B}.\frac{\partial\textbf{B}}{\partial
t})\textbf{v}\nonumber\\
&&+\textbf{V}.(\textbf{B}.\delta\textbf{B})\textbf{V}
+\textbf{V}.(\delta\textbf{B}\frac{\partial\textbf{B}}{\partial
t})\textbf{V}-\textbf{v}.(\textbf{B}.\textbf{V})\frac{\partial\textbf{B}}{\partial
t}-\textbf{V}.(\textbf{B}.\textbf{V})\frac{\partial\delta\textbf{B}}
{\partial t}\nonumber\\
&&-\textbf{V}.(\textbf{B}.\textbf{v})\frac{\partial\delta\textbf{B}}{\partial
t}-\textbf{V}.(\delta\textbf{B}.\textbf{V})\frac{\partial\textbf{B}}{\partial
t}]-\frac{1}{4\pi\alpha}[\textbf{V}.(\textbf{B}.\textbf{B})\frac{\partial\textbf{v}}{\partial
t}-\textbf{V}.(\textbf{B}.\frac{\partial\delta\textbf{v}}{\partial
t})\textbf{B}]\nonumber\\
&&-\frac{1}{4\pi\alpha}\nabla\times(\alpha\delta\textbf{B}).
\end{eqnarray}

The component form of these equations, using Eq.(\ref{11}), can be
composed as follows
\begin{eqnarray}
\label{17}&&\frac{1}{\alpha}\frac{\partial b_x}{\partial
t}-ub_{x,z}=(ub_x-Vb_z-v_x+\lambda v_z)\nabla
\ln\alpha\nonumber\\
&&-(v_{x,z}-\lambda
v_{z,z}-\lambda'v_z+V'b_z+Vb_{z,z}-u'b_x),\\\label{18}
&&\frac{1}{\alpha}\frac{\partial b_z}{\partial t}=0,\\\label{19}
&&b_{z,z}=0,
\end{eqnarray}
\begin{eqnarray}\label{20}
&&\rho\frac{\partial\tilde{\rho}}{\partial
t}+p\frac{\partial\tilde{p}}{\partial
t}+(\rho+p)\gamma^2(V\frac{\partial v_x}{\partial t}+u\frac{\partial
v_z}{\partial t})+\alpha u\rho\rho_{,z}+\alpha
upp_{,z}\nonumber\\
&&+\alpha(\rho+p)\{\gamma^2uVv_{x,z}+(1+\gamma^2u^2)v_{z,z}\}-\frac{1}{\gamma}
(\tilde{\rho}-\tilde{p})(\alpha u\gamma
p)_{,z}\nonumber\\
&&+\alpha(\rho+p)\gamma^2u
\{(1+2\gamma^2V^2)V'+2\gamma^2uVu'\}v_x-\alpha(\rho+p)\nonumber\\
&&\times\{(1-2\gamma^2u^2)(1+\gamma^2u^2)\frac{u'}{u}\}-2\gamma^4u^2VV'\}v_z=0,\\\label{21}
&&\left\{(\rho+p)\gamma^2(1+\gamma^2V^2)
+\frac{B^2}{4\pi}\right\}\frac{1}{\alpha}\frac{\partial
v_x}{\partial t}+\left\{(\rho+p)\gamma^4uV-\frac{\lambda B
^2}{4\pi}\right\}\nonumber\\
&&\times\frac{1}{\alpha}\frac{\partial v_z}{\partial
t}+\left\{(\rho+p)\gamma^2(1+\gamma^2V^2)
+\frac{B^2}{4\pi}\right\}uv_{x,z}+\left\{(\rho+p)\gamma^4uV\right.\nonumber\\
&&\left.-\frac{\lambda B^2}{4\pi}\right\}uv_{z,z}
-\frac{B^2}{4\pi}(1+u^2)b_{x,z}-\frac{B^2}{4\pi\alpha}\left\{\alpha'(1+u^2)+\alpha
uu'\right\}b_x\nonumber\\
&&+\gamma^2u(\rho\tilde{\rho}+p\tilde{p})\left\{(1+\gamma^2V^2)V'+\gamma^2uVu'\right\}
+[(\rho+p)\gamma^4u\{(1\nonumber\\
&&+4\gamma^2V^2)uu'+4VV'(1+\gamma^2V^2)\}+\frac{B^2u\alpha'}{4\pi\alpha}]v_x+[(\rho+p)
\gamma^2\{(1\nonumber\\
&&+2\gamma^2u^2)(1+2\gamma^2V^2)V'-\gamma^2V^2V'
+2\gamma^2(1+2\gamma^2u^2)uVu'\}\nonumber\\
&&-\frac{B^2u} {4\pi\alpha}(\lambda\alpha)']v_z=0,\\\label{22}
&&\left\{(\rho+p)\gamma^2(1+\gamma^2u^2)
+\frac{\lambda^2B^2}{4\pi}\right\}\frac{1}{\alpha}\frac{\partial
v_z}{\partial t}+\left\{(\rho+p)\gamma^4uV -\frac{\lambda B
^2}{4\pi}\right\}\nonumber\\
&&\times\frac{1}{\alpha}\frac{\partial v_x}{\partial t}
+\left\{(\rho+p)\gamma^2(1+\gamma^2u^2)+\frac{\lambda^2B^2}{4\pi}\right\}
uv_{z,z}+\{(\rho+p)\gamma^4u\nonumber\\
&&\times V-\frac{\lambda B^2}{4\pi}\}uv_{x,z}+\frac{\lambda
B^2}{4\pi}(1+u^2)b_{x,z}+\frac{B^2}{4\pi\alpha}\{\alpha'\lambda-(\alpha\lambda)'
+u\lambda\nonumber\\
&&\times(u\alpha'+u'\alpha)\}b_x+(\rho\tilde{\rho}+p\tilde{p})\gamma^2\{a_z
+u u'(1+\gamma^2u^2)+\gamma^2u^2VV'\}\nonumber\\
&&+[(\rho+p)\gamma^4\{u^2V'(1+4\gamma^2V^2)+2V(a_z+uu'(1+2\gamma^2u^2))\}-\lambda
B^2\nonumber\\&&\times\frac{u\alpha'}{4\pi\alpha}]v_x+[(\rho+p)
\gamma^2\{u'(1+\gamma^2u^2)(1+4\gamma^2u^2)+2u\gamma^2(a_z+(1\nonumber\\
&&+2\gamma^2u^2)V V')\}+\frac{\lambda
B^2u}{4\pi\alpha}(\alpha\lambda)']v_z+(p'\tilde{p}+p\tilde{p}')=0,
\end{eqnarray}
\begin{eqnarray}\label{23}
&&\frac{1}{\alpha}\gamma^2\rho\frac{\partial
\tilde{\rho}}{\partial t}+\frac{1}{\alpha}\gamma^2p\frac{\partial
\tilde{p}}{\partial
t}+\gamma^2(\rho'+p')v_z+u\gamma^2(\rho\tilde{\rho}_{,z}+p\tilde{p}_{,z}+\rho'\tilde{\rho}\nonumber\\
&&+p'\tilde{p})-\frac{1}{\alpha}p\frac{\partial
\tilde{p}}{\partial
t}+2\gamma^2u(\rho\tilde{\rho}+p\tilde{p})a_z+\gamma^2u'(\rho\tilde{\rho}+p\tilde{p})+2(\rho\nonumber\\
&&+p)\gamma^4(uV'+2uVa_z+u'V)v_x+2(\rho+p)\gamma^2(2\gamma^2uu'+a_z\gamma^4\nonumber\\
&&+2\gamma^2u^2a_z)v_z+2(\rho+p)\gamma^4uVv_{x,z}+(\rho+p)\gamma^2(1+2\gamma^2u^2)\nonumber\\
&&\times v_{z,z}-\frac{B^2}{4\pi\alpha}[(V^2+u^2)\lambda
b_x+(V^2+u^2)b_z-\lambda V(\lambda V\nonumber\\
&&+u)\frac{\partial b_x}{\partial t}-u(\lambda V+u)\frac{\partial
b_z}{\partial t}]-\frac{B^2}{4\pi\alpha}[(V-\lambda
u)v_{x,t}+\lambda(u\lambda\nonumber\\
&&-V)v_{z,t}]+\frac{B}{4\pi}b_{x,z}=0.
\end{eqnarray}
For Fourier analysis, we take the harmonic spacetime dependence of
perturbation
\begin{eqnarray}\label{24}
\widetilde{\rho}(t,z)=c_1e^{-\iota(\omega t-kz)},&\quad&
\widetilde{p}(t,z)=c_2e^{-\iota(\omega t-kz)},\nonumber\\
v_z(t,z)=c_3e^{-\iota(\omega t-kz)},&\quad&
v_x(t,z)=c_4e^{-\iota(\omega t-kz)},\nonumber\\
b_z(t,z)=c_5e^{-\iota(\omega t-kz)},&\quad&
b_x(t,z)=c_6e^{-\iota(\omega t-kz)}.
\end{eqnarray}
Here $k$ is the $z$-component of the wave vector $(0,0,k)$ and
$\omega$ is the angular frequency. Using the wave vector, we obtain
refractive index which helps to examine the behavior of plasma waves
near the event horizon. Wave vector and dispersion relation can be
defined as
\begin{itemize}
\item\textbf{Wave Vector}: A vector whose direction indicates the
direction of phase propagation of a wave is called wave vector.
Its magnitude is the wave number.
\item\textbf{Dispersion Relation}:  Dispersion relation gives the
angular frequency $\omega$ as a function of wave vector $k$ in the
form $\frac{\omega}{k}=\lambda\nu$, where $\lambda$ is the
wavelength \cite{27}. Dispersion is said to be anomalous if the
refractive index is less than one and its change with respect to
angular frequency is negative, otherwise normal.
\end{itemize}

The Fourier analyzed form of Eqs.(\ref{17})-(\ref{23}), by using
Eq.(\ref{24}), are
\begin{eqnarray}
\label{25}&&c_{4}(\alpha'+\iota k\alpha)-c_3\
\left\{(\alpha\lambda)'+\iota k\alpha\lambda\ \right\}-c_5(\alpha
V)'+c_6\{(\alpha
u)'+\iota\omega\nonumber\\
&&+\iota ku\alpha\}=0,\\\label{26}
&&c_5(\frac{-\iota\omega}{\alpha})=0,\\\label{27} &&c_5\iota
k=0,\\\label{28} &&c_1\{(-\iota\omega+\iota k\alpha
u)\rho-p\gamma^2\alpha u(VV'+u u')-\alpha'up-\alpha u'p-\alpha
up'\}\nonumber\\
&&+c_2\{(-\iota\omega+\iota k\alpha u)p+\alpha'up+\alpha
u'p+\alpha
up'+p\gamma^2\alpha u(VV'+u u')\}\nonumber\\
&&+c_3(\rho+p)[-\iota\omega\gamma^2u+\iota
k\alpha(1+\gamma^2u^2)-\alpha\{(1-2\gamma^2u^2)(1+\gamma^2u^2)
\nonumber\\
&&\times\frac{u'}{u}-2\gamma^4u^2VV'\}]+c_4(\rho+p)[\gamma^2V(-\iota\omega+\iota
k\alpha
u)+\alpha\gamma^2u\{(1\nonumber\\
&&+2\gamma^2V^2)V'+2\gamma^2uVu'\}]=0,\\\label{29}
&&c_1\rho\gamma^2u\{(1+\gamma^2V^2)V'+\gamma^2uVu'\}+c_2p\gamma^2u\{(1+\gamma^2V^2)V'\nonumber\\
&&+\gamma^2uVu'\}+c_3[-\{(\rho+p)\gamma^4uV-\frac{\lambda\
B^2}{4\pi}\}\frac{\iota\omega}{\alpha}+\{(\rho+p)\gamma^4uV\nonumber\\
&&-\frac{\lambda\ B^2}{4\pi}\}\iota
ku+(\rho+p)\gamma^2\{(1+2\gamma^2u^2)(1+2\gamma^2V^2)-\gamma^2V^2\}V'\nonumber\\
&&+2\gamma^4(\rho+p)uVu'(1+2\gamma^2u^2)-\frac{B^2u}{4\pi\alpha}(\alpha\lambda)']+c_4[-\{(\rho+p)
\gamma^2\nonumber\\
&&(1+\gamma^2V^2)+\frac{B^2}{4\pi}\}\frac{\iota\omega}{\alpha}
+\{(\rho+p)\gamma^2(1+\gamma^2V^2)+\frac{B^2}{4\pi}\}\iota ku\nonumber\\
&&+(\rho+p)\gamma^4u\{(1+4\gamma^2V^2)u
u'+4VV'(1+\gamma^2V^2)\}+\frac{B^2u\alpha'}{4\pi\alpha}]\nonumber\\
&&-c_6\frac{B^2}{4\pi}\{(1+u^2)\iota
k+(1+u^2)\frac{\alpha'}{\alpha}+uu'\}=0,\\\label{30}
&&c_1\rho\gamma^2\{a_z+uu'(1+\gamma^2u^2)+\gamma^2u^2VV'\}+c_2[p\gamma^2\{a_z+uu'\nonumber\\
&&(1+\gamma^2u^2)+\gamma^2u^2VV'\}+p'+\iota k
p]+c_3[-\{(\rho+p)\gamma^2(1+\gamma^2u^2)\nonumber\\
&&+\frac{\lambda
^2B^2}{4\pi}\}\frac{\iota\omega}{\alpha}+\{(\rho+p)\gamma^2(1+\gamma^2u^2)+\frac{\lambda
^2B^2}{4\pi}\}\iota ku+\{(\rho+p)\gamma^2\nonumber\\
&&\times \{u'(1+\gamma^2u^2)(1+4\gamma^2u^2)+2u\gamma^2\{a_z+(1+2\gamma^2u^2)\}VV'\}\nonumber\\
&&+\frac{\lambda
B^2u}{4\pi\alpha}(\alpha\lambda)']+c_4[-\{(\rho+p)\gamma^4u
V-\frac{\lambda
B^2}{4\pi}\}\frac{\iota\omega}{\alpha}+\{(\rho+p)\gamma^4u
V\nonumber
\end{eqnarray}
\begin{eqnarray}
&&-\frac{\lambda B^2}{4\pi}\}\iota
ku+\{(\rho+p)\gamma^4\{u^2V'(1+4\gamma^2V^2)+2V
\{(1+2\gamma^2u^2)uu'\nonumber\\
&&+a_z-\frac{\lambda B^2\alpha'
u}{4\Pi\alpha}\}]+c_6[\frac{B^2}{4\Pi\alpha}\{-(\alpha\lambda)'
+\alpha'\lambda-u\lambda(u\alpha'+u'\alpha)\}\nonumber\\
&&+\frac{\lambda B^2}{4\Pi}(1+u^2)\iota k]=0,\\
\label{31}&&c_1\{(\frac{-\iota\omega}{\alpha}\gamma^2+\iota ku
\gamma^2+2u\gamma^2a_z+\gamma^2u')\rho+u\rho'\gamma^2\}
+c_2\{(\frac{\iota\omega}{\alpha}(1-\gamma^2)\nonumber\\
&&+\iota ku\gamma^2+2\gamma^2ua_z+\gamma^2u')p+u\gamma^2p'\}+c_3\gamma^2\{(\rho'+p')+2\nonumber\\
&&\times(2\gamma^4uu'+a_z+2\gamma^2u^2a_z)(\rho+p)+(1+2\gamma^2u^2)(\rho+p)\iota
k+\frac{\lambda B^2}{4\pi\alpha}\nonumber\\
&&\times(\lambda u-V)\iota\omega\}+c_4[2(\rho+p)\gamma^2\{(u
V'+2uVa_z+u'V)+uV\iota
k\}\nonumber\\
&&+\frac{B^2}{4\pi\alpha}(V-u\lambda)\iota\omega]
+c_6[\frac{-B^2}{4\pi\alpha}\{(V^2+u^2)\lambda+\lambda
V(\lambda V+u)\iota\omega\}\nonumber\\
&&+\frac{B}{4\Pi}\iota k]=0.
\end{eqnarray}
These equations will be used to find dispersion relations.

\section{Plasma Flow with Rotating Non-Magnetized Background}

In the present section we shall study the rotating non-magnetized
plasma, i.e., $\textbf{B}=0$. Equations (\ref{4}) and (\ref{5})
which are evolution equations of the magnetic field vanish. In the
Fourier analyzed perturbed GRMHD equations
(Eqs.(\ref{28})-(\ref{31})), we put $B=0=\lambda$ and $c_5=0=c_6$
and obtain
\begin{eqnarray}{\setcounter{equation}{1}}
\label{32}&&c_1(\frac{-\iota\omega}{\alpha}\rho)+c_2(\frac{-\iota\omega}{\alpha}p)
+c_3(\rho+p)[\frac{-\iota\omega}{\alpha}\gamma^2u+(1+\gamma^2u^2)\iota k\nonumber\\
&&-(1-2\gamma^2u^2)(1+\gamma^2u^2)\frac{u'}{u}+2\gamma^4u^2VV']
+c_4(\rho+p)\gamma^2[(\frac{-\iota\omega}{\alpha}\nonumber\\
&&+\iota ku)V +u(1+2\gamma^2V^2)V'+2\gamma^2u^2Vu']=0,\\\label{33}
&&c_1\rho\gamma^2u\{(1+\gamma^2V^2)V'+\gamma^2Vuu'\}+c_2p\gamma^2u
\{(1+\gamma^2V^2)V'\nonumber\\
&&+\gamma^2Vuu'\}+c_3(\rho+p)\gamma^2
\{(\frac{-\iota\omega}{\alpha}+\iota ku)\gamma^2V
u+(1+2\gamma^2u^2)\nonumber\\
&&\times(1+2\gamma^2V^2)V'-\gamma^2V^2V'
+2\gamma^2(1+2\gamma^2u^2) uVu'\}+c_4(\rho\nonumber
\end{eqnarray}
\begin{eqnarray}
&&+p)\gamma^2[(1+\gamma^2V^2)(\frac{-\iota\omega}{\alpha}+\iota
ku)+\gamma^2u\{(1+4\gamma^2u^2)u u'+4V\nonumber\\
&&V'(1+\gamma^2V^2)\}]=0,\\\label{34}
&&c_1\rho\gamma^2\{a_z+(1+\gamma^2u^2)u
u'+\gamma^2u^2VV'\} +c_2[p\gamma^2\{a_z+(1+\gamma^2u^2)\nonumber\\
&&\times u u'+\gamma^2u^2VV'\}+(p'+\iota k
p)]+c_3(\rho+p)\gamma^2[(1+\gamma^2u^2)(\frac{-\iota\omega}{\alpha}\nonumber\\
&&+\iota ku)+u'(1+\gamma^2u^2)
(1+4\gamma^2u^2)+2u\gamma^2\{a_z+(1+2\gamma^2u^2)\nonumber\\
&&\times V V'\}]+c_4(\rho+p)\gamma^4[u
V(\frac{-\iota\omega}{\alpha}+
\iota ku)+u^2V'(1+4\gamma^2u^2)\nonumber\\
&&+2V\{a_z+(1+2\gamma^2u^2)u u'\}]=0,\\\label{35}
&&c_1\{(\frac{-\iota\omega}{\alpha}\gamma^2+\iota ku
\gamma^2+2u\gamma^2a_z+\gamma^2u')\rho+u\rho'\gamma^2\}
+c_2\{(\frac{\iota\omega}{\alpha}(1-\gamma^2)\nonumber\\
&&+\iota
ku\gamma^2+2\gamma^2ua_z+\gamma^2u')p+u\gamma^2p'\}+c_3\gamma^2\{(\rho'+p')+2
(2\gamma^2uu'\nonumber\\
&&+a_z+2\gamma^2u^2a_z)(\rho+p)+(1+2\gamma^2u^2)(\rho+p)\iota
k\}+c_4(\rho+p)2\gamma^4\nonumber\\
&&\{(u V'+2uVa_z+u'V)+uV\iota k\}=0.
\end{eqnarray}

\subsection{Numerical Solutions}

In order to determine numerical solutions, we use the the following
assumptions
\begin{itemize}
\item Specific enthalpy: $\mu=1$.
\item Time lapse: $\alpha=\tanh(10z)/10,$
\item Stationary fluid: $\alpha\gamma=1$ with velocity components $V=u$
yields the following expression:
$\alpha\gamma=1~\Rightarrow\gamma=1/\sqrt{1-u^2-V^2}=1/\alpha$.
\item  Velocity components: $u=V,~x$ and $z$-components of velocity
lead to $u=V=-\sqrt\frac{1-\alpha^2}{2}$.
\end{itemize}
With the assumption of stiff fluid, i.e., $\rho=p$, when we
replace these values, the mass conservation law in three
dimensions yields $\rho=p=-1/2u$. The GRMHD equations
(Eqs.(\ref{4})-(\ref{8})) are satisfied by the above assumptions
for the region $1.5\leq z\leq10, 0\leq\omega\leq 10$. For plasma
flow in rotating non-magnetized background, we acquire the
constant values of $u=V=-0.703562$ which make the flow constants
$l=0.703562$ and $e=1$. The determinant of the coefficients of
constants of Eqs.(\ref{32})-(\ref{35}) are solved which lead to a
complex dispersion relation \cite{28}. By comparison of real and
imaginary parts, two dispersion relations are found. The real part
of the determinant gives an equation quartic in $k$
\begin{equation}\label{36}
A_1(z)k^4+A_2(z,\omega)k^3+A_3(z,\omega)k^2+A_4(z,\omega)k+A_5(z,\omega)=0
\end{equation}
which gives two real values of $k$ and two complex values
conjugate. The equation obtained from the complex part is cubic in
$k$
\begin{equation}\label{37}
B_1(z)k^3+B_2(z,\omega)k^2+B_3(z,\omega)k+B_4(z,\omega)=0
\end{equation}
which yields one real value of $k$ and the remaining two are
complex conjugate of each other. We can compute refractive index
and its change with respect to angular frequency by using the real
values obtained from Eq.(\ref{36}) and (\ref{37}). The results
reached are shown in Figures $1,~2$ and $3$. The following table
shows the results obtained from these
Figures.\\
\begin{center}
Table I. Direction and refractive index of waves.
\end{center}
\begin{tabular}{|c|c|c|c|c|}
\hline & \textbf{Direction of Waves} & \textbf{Refractive Index}
($n$)\\ \hline
& & $n<1$ in the region \\
\textbf{1} & Move towards the event horizon & $0\leq z\leq
10,8.5\leq\omega\leq 10$ and\\&& $0\leq z\leq 10,9.5\leq\omega\leq
10$\\&&with the decrease in $z$  \\
\hline
& & $n<1$ in the region\\
\textbf{2} & Move towards the event horizon & $2\leq z\leq
10,2\leq\omega\leq 9$ and\\&& $2\leq z\leq 10,9.9\leq\omega\leq
10$\\&&with the decrease in $z$  \\
\hline
& & $n<1$ in the region\\
\textbf{3} & Move away from the event horizon & $0\leq z\leq10,2\leq\omega\leq 10$\\
& &with the decrease in $z$ \\
\hline
\end{tabular}
\\\\

In Figure \textbf{1}, the change in refractive index with respect
to angular frequency represents normal as well as anomalous
dispersion of waves at random points. In Figure \textbf{2},
dispersion is anomalous in the regions $2\leq z\leq 10,
2.2\leq\omega\leq 2.25$ and $0\leq z\leq 9, 7.5\leq\omega\leq 8$
while Figure \textbf{3} also shows normal and anomalous dispersion
randomly. The group and phase velocities are found to be
antiparallel in all figures.
\begin{figure}
\center \epsfig{file=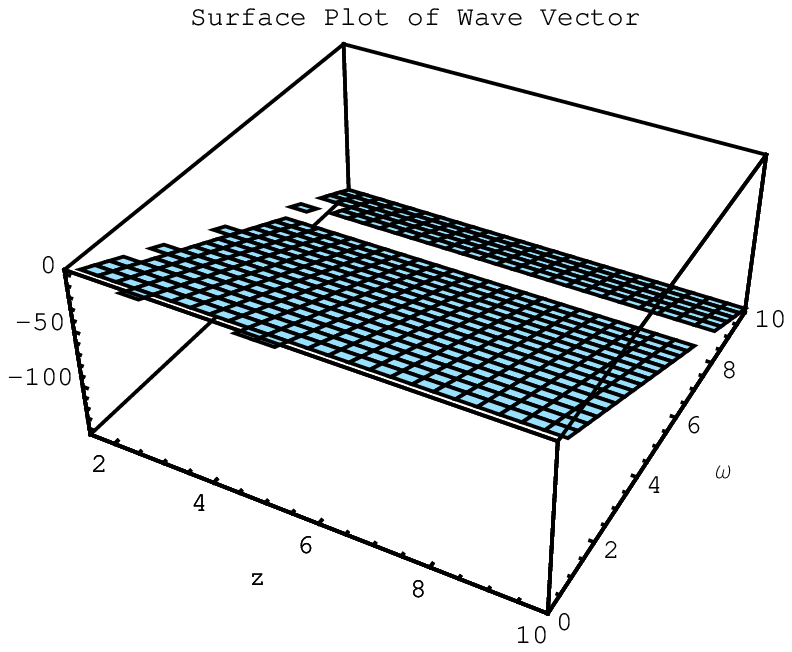,width=0.40\linewidth} \center
\begin{tabular}{cc}
\epsfig{file=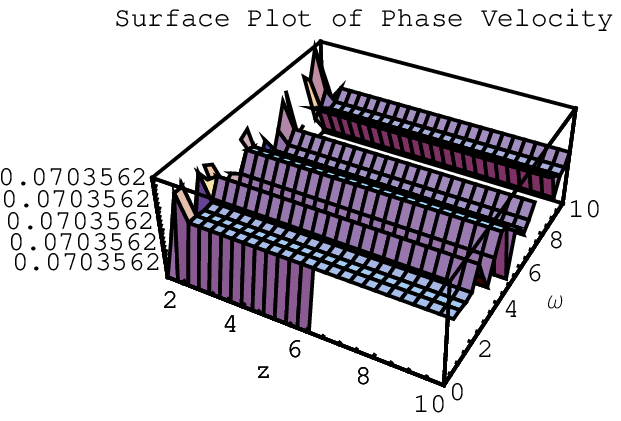,width=0.45\linewidth}
\epsfig{file=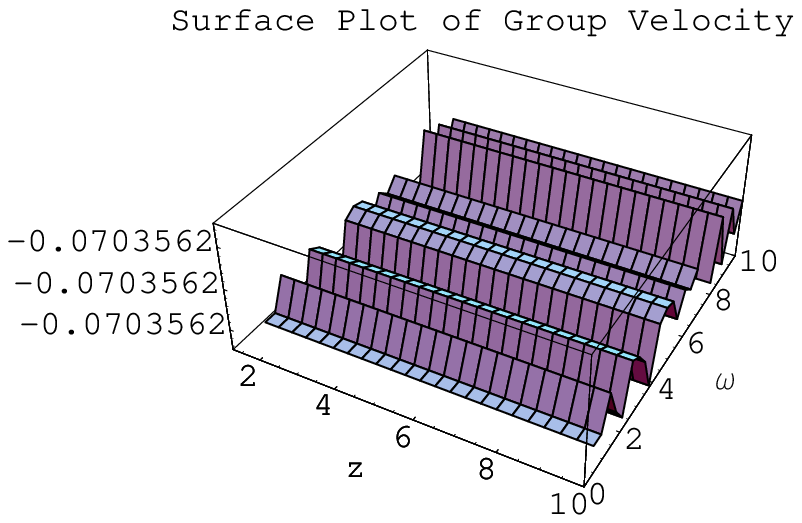,width=0.45\linewidth}\\
\epsfig{file=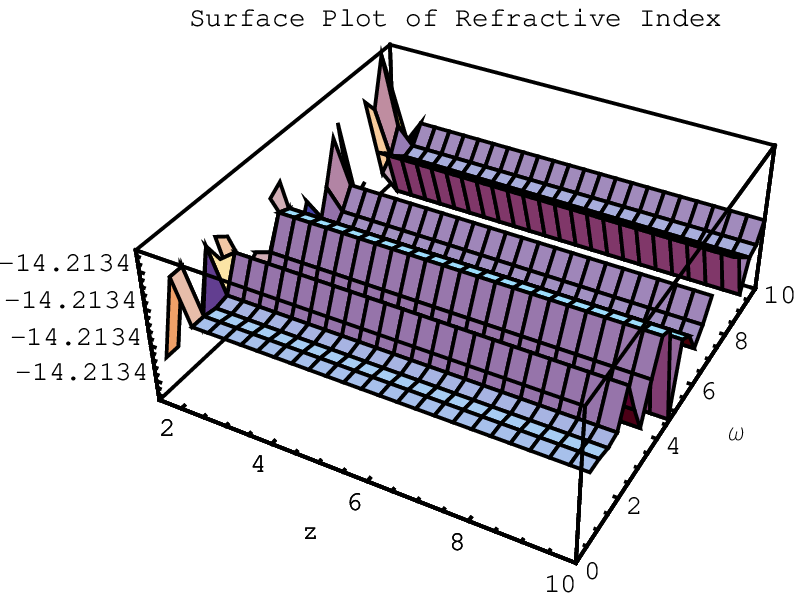,width=0.40\linewidth}
\epsfig{file=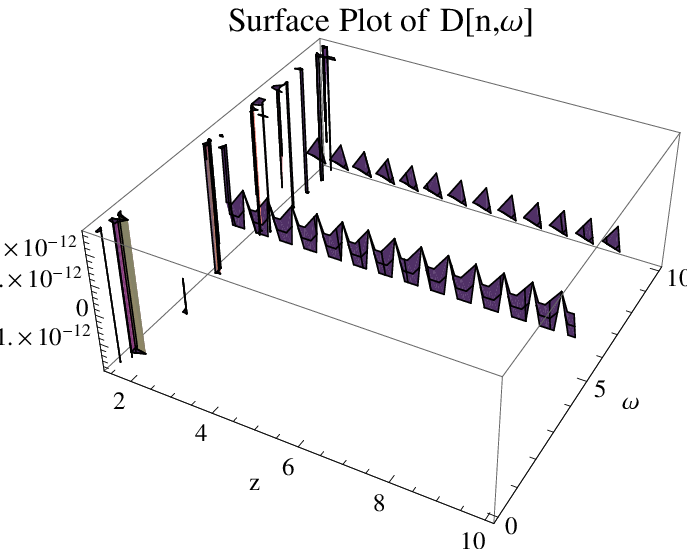,width=0.35\linewidth}
\end{tabular}
\caption{Waves move towards the event horizon. The dispersion is
normal as well as anomalous at random points.}
\end{figure}

\begin{figure} \center
\epsfig{file=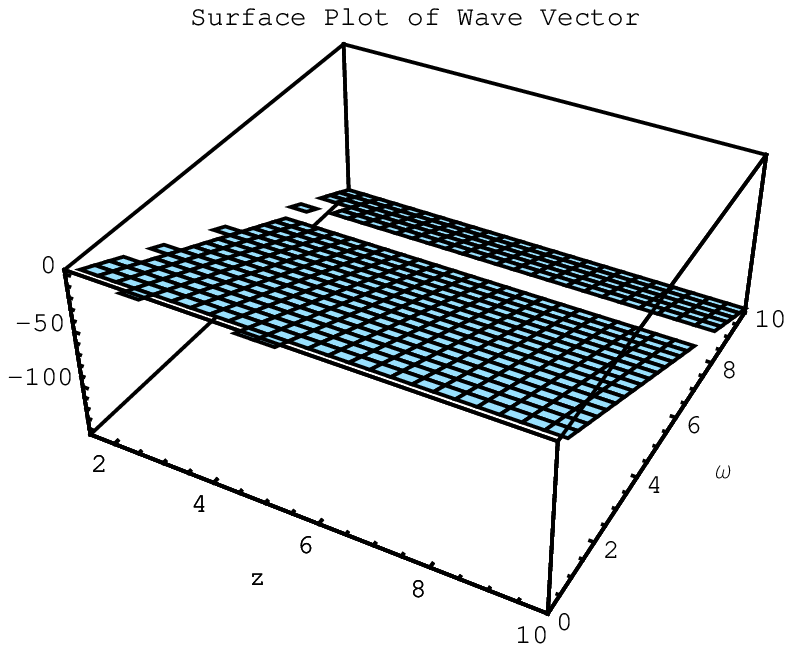,width=0.40\linewidth} \center
\begin{tabular}{cc}
\epsfig{file=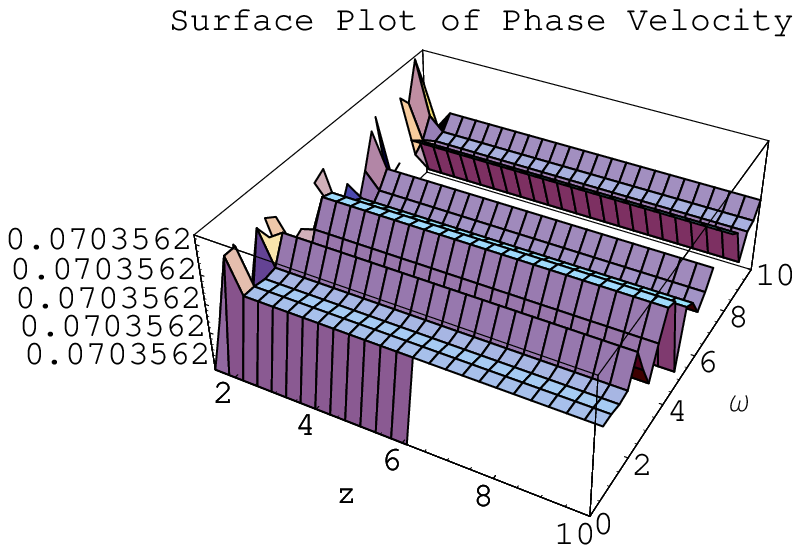,width=0.45\linewidth}
\epsfig{file=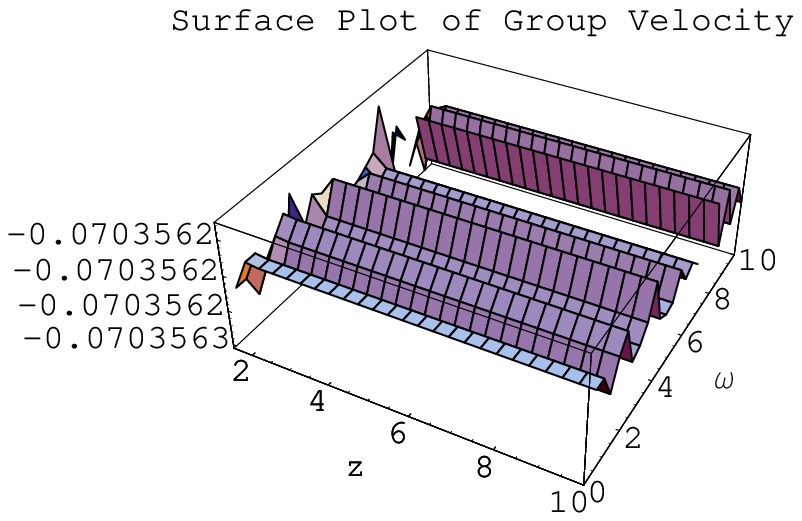,width=0.45\linewidth}\\
\epsfig{file=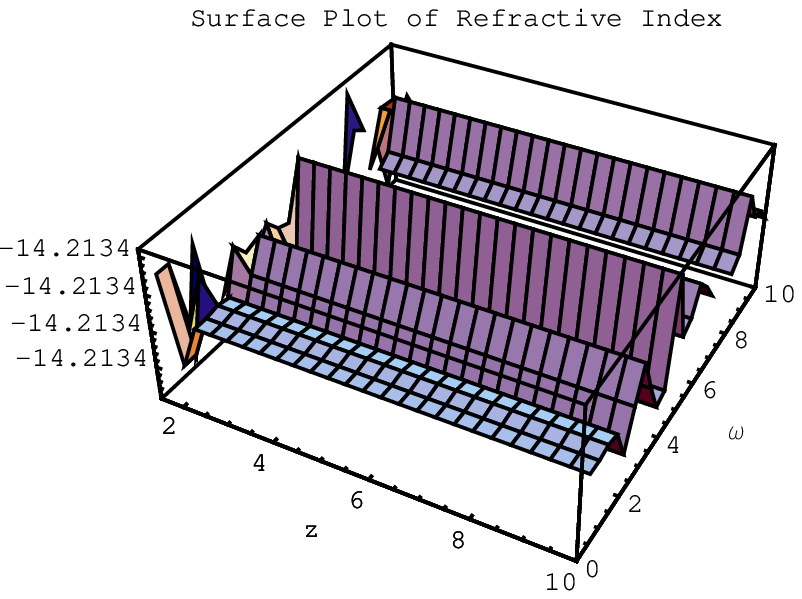,width=0.40\linewidth}
\epsfig{file=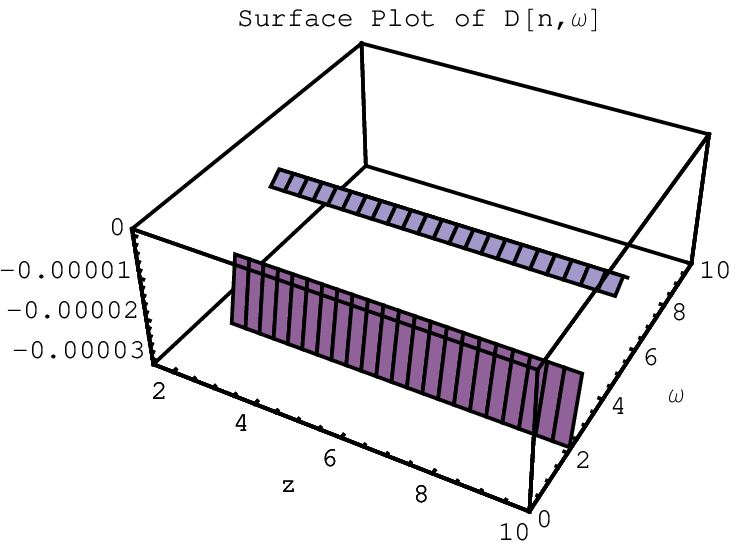,width=0.40\linewidth}
\end{tabular}
\caption{Waves are directed towards the event horizon. Region has
anomalous dispersion.}
\end{figure}

\begin{figure}
\center \epsfig{file=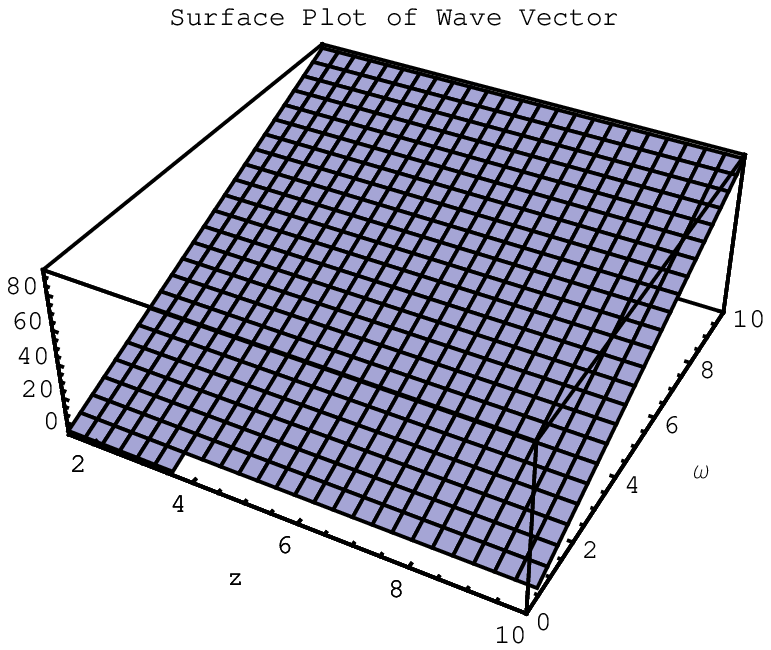,width=0.40\linewidth} \center
\begin{tabular}{cc}
\epsfig{file=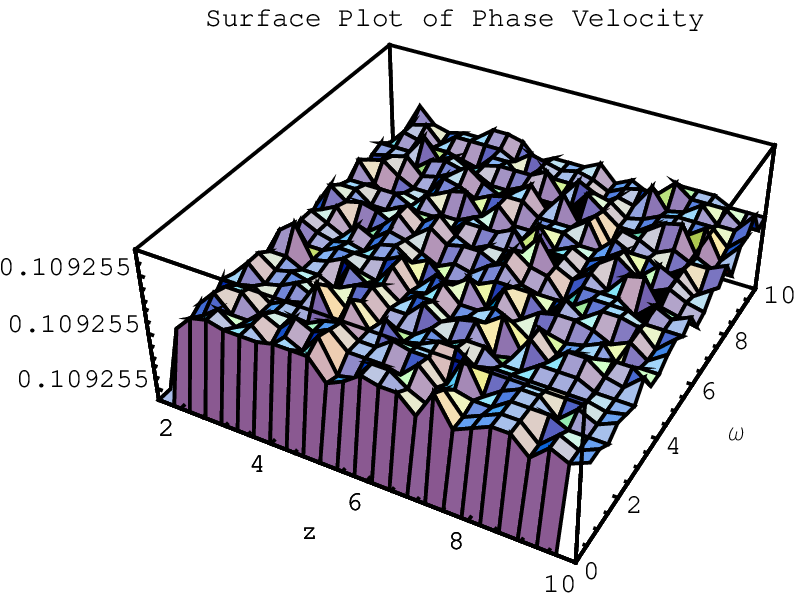,width=0.40\linewidth}
\epsfig{file=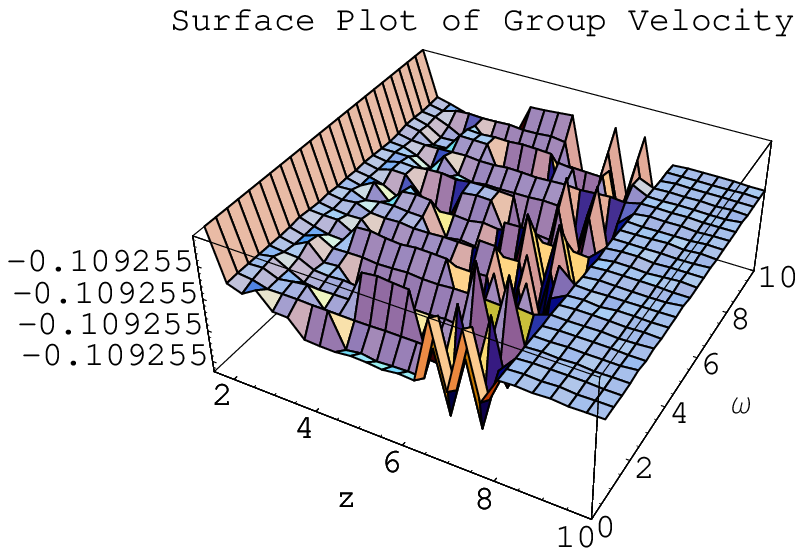,width=0.45\linewidth}\\
\epsfig{file=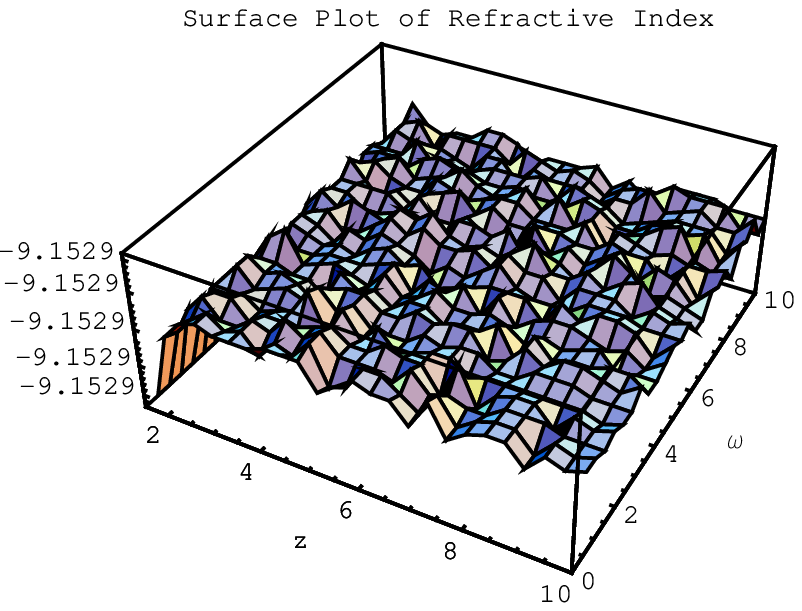,width=0.40\linewidth}
\epsfig{file=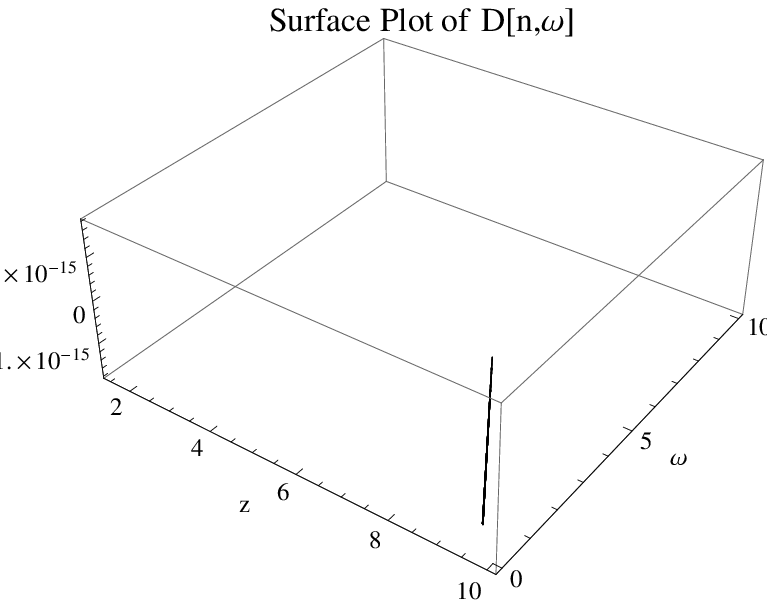,width=0.35\linewidth}
\end{tabular}
\caption{Waves are directed away from the event horizon. The region
has normal and anomalous dispersion of waves.}
\end{figure}

\section{Plasma Flow with Rotating Magnetized Background}

This is the general case in which we assume that plasma is
magnetized and rotating. We suppose $xz$-plane for the velocity and
magnetic field of fluid. Equations (\ref{25})-(\ref{31}) are the
corresponding Fourier analyzed perturbed GRMHD equations.

\subsection{Numerical Solutions}

For magnetized background, we assume $V^F=0$ with $u=V$ so that
Eq.(\ref{a}) gives $\lambda=1$. Also, we consider
$B=\sqrt{\frac{176}{7}}$ which leads to $\frac{B^2}{4\pi}=2$.
Further, the values of lapse function, velocity, pressure, density
and specific enthalpy remain the same as given in section 4.1. The
above assumed values satisfy the perfect GRMHD
Eqs.(\ref{4})-(\ref{8}) for the range $1.5\leq z\leq10,~
0\leq\omega\leq 10$. The values of flow constants are
$f=e=1,~l=-7.32001$ and $h=-1.76321$. Also, we have $u=V=-0.703562$.
We obtain the determinant of the coefficients of constants by
substituting these values in Eqs.(\ref{25}) and
(\ref{28})-(\ref{31}) which gives two dispersion relations. Also,
$c_5=0$ from Eqs.(\ref{26})-(\ref{27}). The dispersion relation
obtained from the real part has the following form
\begin{equation}{\setcounter{equation}{1}}\label{38}
A_1(z)k^4+A_2(z,\omega)k^3+A_3(z,\omega)k^2+A_4(z,\omega)k+A_5(z,\omega)=0
\end{equation}
while the imaginary part gives the following dispersion relation
\begin{eqnarray}\label{39}
&&B_1(z)k^5+B_2(z,\omega)k^4+B_3(z,\omega)k^3+B_4(z,\omega)k^2+B_5(z,\omega)k\nonumber\\
&&+B_6(z,\omega)=0.
\end{eqnarray}
We have used the software \emph{Mathematica} to solve the above
equations so that the roots can be displayed in terms of graphs.
However, we could not find any graph of the roots. It seems that
either all the roots are imaginary or there does not exist any wave
in this region.

\section{Summary}

This paper investigates the isothermal plasma wave properties of the
Schwarzschild magnetosphere in a Veselago medium. For this purpose,
we have re-formulated the Maxwell and $3+1$ GRMHD equations by
considering both permittivity and permeability less than zero. The
component and Fourier analysi of these equations are derived for
rotating plasma in non-magnetized and magnetized backgrounds.

For the rotating non-magnetized plasma, our assumed values satisfy
the $3+1$ GRMHD equations in the region $1.5\leq z \leq10$. In
Figures \textbf{1} and \textbf{2}, waves move towards the event
horizon. The dispersion is normal as well as anomalous at random
points in the first figure while it is anomalous in the second
figure. The third figure indicates that waves move towards the event
horizon while it has normal and anomalous dispersion of waves
randomly. All the figures show that phase and group velocities are
antiparallel. Refractive index is less than one and it increases in
a small region near the event horizon. From the previous literature
\cite{19} of the usual medium for isothermal plasma, we know that
all waves move towards the event horizon. In the Veselago medium,
Figure \textbf{3} indicates that waves move away from the event
horizon. This is entirely a different result in this unusual medium.
For the usual medium, the value of refractive index is always
greater than one while it is less than one in all figures. This
confirms the presence of Veselago medium. The rotating magnetized
case does not provide any explicit graph which indicates that waves
may not be found in the region $1.5\leq z \leq10$.

It would be interesting to extend this work by assuming hot plasma
in this unusual medium. The Kerr spacetime can also be used to
analyze the wave properties in this unusual medium.

\renewcommand{\theequation}{A\arabic{equation}}

\section*{Appendix A}

In Veselago medium ($\epsilon<0, \mu<0$), Maxwell equations, the
GRMHD equations for the general line element and the Schwarzschild
planar analogue are given in this appendix. Maxwell equations in
this unusual medium are
\begin{eqnarray}{\setcounter{equation}{1}}
\label{40}&&\nabla.\textbf{B}=0,\\
\label{41}&&\nabla\times\textbf{E}+\frac{\partial\textbf{B}}{\partial
t}=0,\\
\label{42}&&\nabla\cdot\textbf{E}=-\frac{\rho_e}{\epsilon},\\
\label{43}&&\nabla\times\textbf{B}=-\mu\textbf{j}+\frac{\partial\textbf{E}}{\partial
t}=0.
\end{eqnarray}
The GRMHD equations in this medium will be
\begin{eqnarray}\label{44}
&&\frac{d\textbf{B}}{d\tau}
+\frac{1}{\alpha}(\textbf{B}.\nabla)\beta
+\theta\textbf{B}=-\frac{1}{\alpha}\nabla\times(\alpha\textbf{V}\times\textbf{B}),\\
\label{45}
&&\nabla.\textbf{B}=0,\\
\label{46}
&&\frac{D\rho_0}{D\tau}+\rho_0\gamma^2\textbf{V}.\frac{D\textbf{V}}{D\tau}
+\frac{\rho_0}{\alpha}\left\{\frac{g,_t}{2g}+\nabla.(\alpha\textbf{V}-\beta)\right\}=0,\\\label{47}
&&\left\{\left(\rho_0\mu\gamma^2+\frac{\textbf{B}^2}{4\pi}\right)\gamma_{ij}
+\rho_0\mu\gamma^4V_iV_j
-\frac{1}{4\pi}B_iB_j\right\}\frac{DV^j}{D\tau}
\nonumber\\
&&+\rho_0\gamma^2V_i\frac{D\mu}{D\tau}
-\left(\frac{\textbf{B}^2}{4\pi}\gamma_{ij}-\frac{1}{4\pi}B_iB_j\right)
V^j_{|k}V^k=-\rho_0\gamma^2\mu\{a_i\nonumber\\
&&-\frac{1}{\alpha}\beta_{j|i}V^j -(\pounds_t\gamma_{ij})V^j\}
-p_{|i}+\frac{1}{4\pi}(\textbf{V}\times
\textbf{B})_i\nabla.(\textbf{V}\times\textbf{B})\nonumber\\
&&-\frac{1}{8\pi\alpha^2}(\alpha\textbf{B})^2_{|i}+\frac{1}{4\pi\alpha}(\alpha
B_i)_{|j}B^j-\frac{1}{4\pi\alpha}(\textbf{B}\times\{\textbf{V}\times
[\nabla\nonumber\\
&&\times(\alpha\textbf{V}\times\textbf{B})
-(\textbf{B}.\nabla)\beta]+(\textbf{V}\times\textbf{B}).\nabla\beta\})_i,\\
\label{48}&&\frac{D}{D\tau}(\mu\rho_0\gamma^2)-\frac{d
p}{d\tau}+\Theta(\mu\rho_0\gamma^2-p)+\frac{1}{2\alpha}
(\mu\rho_0\gamma^2V^{i}V^{j}\nonumber\\&&+p\gamma_{ij})
\pounds_t\gamma_{ij}+2\mu\rho_0\gamma^2(\textbf{V}.\textbf{a})+\mu\rho_0\gamma^2(\nabla.\textbf{V})
-\frac{1}{\alpha}\beta^{j,i}\nonumber\\
&&\times(\mu\rho_0\gamma^2V_{i}V_{j}+p\gamma_{ij})-\frac{1}{4\pi}
(\textbf{V}\times\textbf{B}).(\textbf{V}\times\frac{d\textbf{B}}{d\tau})
-\frac{1}{4\pi}\nonumber\\ &&\times
(\textbf{V}\times\textbf{B}).(\textbf{B}\times\frac{d\textbf{V}}{d\tau})-\frac{1}{4\pi\alpha}
(\textbf{V}\times\textbf{B}).\nabla\beta-\frac{1}{4\pi}\theta(\textbf{V}\times\textbf{B})
\nonumber\\&&+\frac{1}{4\pi\alpha}(\nabla\times\alpha\textbf{}B)=0.
\end{eqnarray}
Since $\beta,~\theta$ and $\pounds_t\gamma_{ij}$ vanish for the
Schwarzschild planar analogue, the perfect GRMHD equations reduce to
\begin{eqnarray}\label{49}
&&\frac{\partial\textbf{B}}{\partial t}=-\nabla \times(\alpha
\textbf{V}\times\textbf{B}),\\\label{50}
&&\nabla.\textbf{B}=0,\\\label{51}
&&\frac{\partial\rho_0}{\partial
t}+(\alpha\textbf{V}.\nabla)\rho_0+\rho_0\gamma^2
\textbf{V}.\frac{\partial\textbf{V}}{\partial
t}+\rho_0\gamma^2\textbf{V}.(\alpha\textbf{V}.\nabla)\textbf{V}\nonumber\\
&&+\rho_0{\nabla.(\alpha\textbf{V})}=0, \\\label{52}
&&\{(\rho_0\mu\gamma^2+\frac{\textbf{B}^2}{4\pi})\delta_{ij}
+\rho_0\mu\gamma^4V_iV_j
-\frac{1}{4\pi}B_iB_j\}(\frac{1}{\alpha}\frac{\partial}{\partial
t}+\textbf{V}.\nabla)V^j\nonumber
\end{eqnarray}
\begin{eqnarray}
&&-(\frac{\textbf{B}^2}{4\pi}\delta_{ij}-\frac{1}{4\pi}B_iB_j)
V^j,_kV^k+\rho_0\gamma^2V_i\{\frac{1}{\alpha}\frac{\partial
\mu}{\partial t}+(\textbf{V}.\nabla)\mu\}\nonumber\\
&&=-\rho_0\mu\gamma^2a_i-p,_i+
\frac{1}{4\pi}(\textbf{V}\times\textbf{B})_i\nabla.(\textbf{V}\times\textbf{B})
-\frac{1}{8\pi\alpha^2}(\alpha\textbf{B})^2,_i\nonumber\\
&&+\frac{1}{4\pi\alpha}(\alpha B_i),_jB^j-\frac{1}{4\pi\alpha}
[\textbf{B}\times\{\textbf{V}\times(\nabla\times(\alpha\textbf{V}
\times\textbf{B}))\}]_i,\\
\label{53}&&(\frac{1}{\alpha}\frac{\partial}{\partial
t}+\textbf{V}.\nabla)(\mu\rho_0\gamma^2)-\frac{1}{\alpha}\frac{\partial
p }{\partial
t}+2\mu\rho_0\gamma^2(\textbf{V}.\textbf{a})+\mu\rho_0\gamma^2
(\nabla.\textbf{V})\nonumber\\&&-\frac{1}{4\pi}
(\textbf{V}\times\textbf{B}).(\textbf{V}\times\frac{1}{\alpha}\frac{\partial
\textbf{B}}{\partial t})-\frac{1}{4\pi}
(\textbf{V}\times\textbf{B}).(\textbf{B}\times\frac{1}{\alpha}\frac{\partial
\textbf{V}}{\partial
t})\nonumber\\&&+\frac{1}{4\pi\alpha}(\nabla\times\alpha\textbf{B})=0.
\end{eqnarray}


\begin{thebibliography}{99}

\bibitem{1} Regge, T. and Wheeler, J.A.: Phy. Rev.
\textbf{108}(1957)1063.

\bibitem{2} Price, R.H.: Phys. Rev. \textbf{D5}, 2419 (1972); ibid. 2439.

\bibitem{3} Sakai, J. and Kawata, T.: J. Phys. Soc. \textbf{49}, 747 (1980).

\bibitem{4} Southwood, D.J. and Kivelson, M.G.: J. Geog. Res.
\textbf{92}, 109 (1987).

\bibitem{5} Gleiser, R.J. Nicasio, C.O. Price, R.H. and Pullin,
J.: Class. Quantum Grav. \textbf{13}, 117 (1996).

\bibitem{6} Arnowitt, R., Deser, S. and Misner, C.W.: Gravitation: An
Introduction to Current Research. John Wiley (1962).

\bibitem{7} Wheeler, J.A.: Battelle Rencontres:
1967 Lectures in Mathematics and Physics, eds. DeWitt, C. and
Wheeler, J.A. W.A. Benjamin Inc. (1968).

\bibitem{8} Macdonald, D.A. and Suen, W.-M.: Phys. Rev.
\textbf{D32}, 848 (1985).

\bibitem{9} Thorne, K.S. and Macdonald, D.A.: Mon. Not. R. Astron.
Soc. \textbf{198}, 339 (1982), ibid. 345.

\bibitem{10} Black Hole: The Membrane Paradigm.
eds. Thorne, K.S. Price, R.H. and Macdonald, D.A. Yale University
Press (1986).

\bibitem{11} Holcomb, K.A. and Tajima, T.: Phys. Rev. \textbf{D40}, 3809 (1989).

\bibitem{12} Holcomb, K.A.: Astrophys. J. \textbf{362}, 381 (1990).

\bibitem{13} Dettmann, C.P., Frankel, N.E. and Kowalenke, V.: Phys.
Rev. \textbf{D48}, 5655 (1993).

\bibitem{14} Buzzi, V., Hines, K.C. and Treumann, R.A.: Phys. Rev.
\textbf{D51}, 6663 (1995), ibid. 6677.

\bibitem{15} Zhang, X.-H.: Phys. Rev. \textbf{D39}, 2933 (1989).

\bibitem{16} Zhang, X.-H.: Phys. Rev. \textbf{D40}, 3858 (1989).

\bibitem{17} Sharif, M. and Sheikh, U.: Gen. Relativ. Gravit.
\textbf{39}, 1437 (2007); ibid. 2095; Int. J. Mod. Phys.
\textbf{A23}, 1417 (2008); J. Korean Phys. Soc. \textbf{52}, 152
(2008); ibid. \textbf{53}, 2198 (2008).

\bibitem{18} Sharif, M. and Sheikh, U.: Class. Quantum Grav.
\textbf{24}, 5495 (2007); Canadian J. Phys. \textbf{87}, 879 (2009);
J. Korean Physical Society \textbf{55}, 1677 (2009).

\bibitem{19} Sharif, M. and Mustafa, G.: Canadian J. Phys. \textbf{86}, 1265 (2008).

\bibitem{20} Sharif, M. and Rafique, A.: Astrophys. Space Sci.
\textbf{325}, 227 (2010).

\bibitem{21} Ziolkowski, R.W. and Heyman, E.: Phys. Rev. E.
\textbf{64}, 056625 (2001).

\bibitem{22} Valanju, P.M., Walser, R.M. and Valanju, A.P.: Phys.
Rev. Lett. \textbf{88}, 187401 (2002).

\bibitem{23} Ross, B.M., Mackay, T.G. and Lakhtakia, A.:
Effect of Charge on NPV Propagation of Electromagnetic Waves in the
Ergosphere of a Rotating Black Hole; astro-ph/0608412.

\bibitem{24} Mackay, T.G. and Lakhtakia, A.: Current Science \textbf{90}, 641 (2006).

\bibitem{25} Krowne, C.M. and Zhang, Y.: Physics of
Negative Refraction and Negative Index Materials. Springer
(2007)p183.

\bibitem{26} Veselago, V.G.: Physics-Uspekhi \textbf{52}, 6 (2009).

\bibitem{27} Crawford Jr., F.S.: Waves. Education Developement Center,
Inc. (1968).

\bibitem{28} Das, A.C.: Space Plasma Physics: An
Introduction. Narosa Publishing House (2004).

\end{thebibliography}
\end{document}